\documentclass[conference]{IEEEtran}
\usepackage{multicol}
\usepackage{lipsum,graphicx}
\usepackage{float}
\usepackage{epstopdf}
\usepackage{ifthen}
\usepackage[latin1]{inputenc}

\usepackage{amssymb}
\usepackage[cmex10]{amsmath}
\usepackage{dcolumn}
\usepackage[nolist]{acronym}

\DeclareGraphicsExtensions{.pdf,.jpeg,.png,.eps}
\usepackage{booktabs}

\usepackage[caption=false,font=footnotesize]{subfig}
\usepackage{comment}
\usepackage{color}
\usepackage[table]{xcolor}
\usepackage{cite}

\usepackage{setspace}
\usepackage{mathrsfs}          
\usepackage{amsmath, amssymb, amsthm}
\usepackage{amsfonts}
\usepackage{mathrsfs}
\usepackage{multirow}        
\usepackage{framed}
\usepackage{psfrag}
\usepackage{graphicx}
\usepackage{textcomp, gensymb} 
\usepackage{algpseudocode} 

\newcolumntype{d}[1]{D{.}{\cdot}{#1} }

\usepackage{bm}
\usepackage{mathtools}
\usepackage{siunitx}

\newcommand{\PreserveBackslash}[1]{\let\temp=\\#1\let\\=\temp}
\newcolumntype{C}[1]{>{\PreserveBackslash\centering}p{#1}}
\newcolumntype{R}[1]{>{\PreserveBackslash\raggedleft}p{#1}}
\newcolumntype{L}[1]{>{\PreserveBackslash\raggedright}p{#1}}

\newcommand{\figref}[1]{Fig.~\ref{#1}}

\newcommand{\tabref}[1]{Table~\ref{#1}}

\usepackage{lipsum}
\usepackage{mathtools}
\usepackage{cuted}
\usepackage{stmaryrd}

\usepackage{tabularx}
\usepackage{tabulary}
\usepackage{tikz}
\usepackage{pgfplots}
\usepgfplotslibrary{units}
\pgfplotsset{compat=1.18}

\usepackage{algpseudocode}
\usepackage{algorithm}

\begin{document}
	\begin{acronym}
\acro{1G}{first generation}
\acro{2G}{second generation}
\acro{3G}{third generation}
\acro{3GPP}{Third Generation Partnership Project}
\acro{4G}{fourth generation}
\acro{5G}{fifth generation}
\acro{802.11}{IEEE 802.11 specifications}
\acro{A/D}{analog-to-digital}
\acro{ADC}{analog-to-digital}
\acro{AGC}{automatic gain control}
\acro{AM}{amplitude modulation}
\acro{AoA}{angle of arrival}
\acro{AP}{access point}
\acro{AR}{augmented reality}
\acro{ASIC}{application-specific integrated circuit}
\acro{ASIP}{Application Specific Integrated Processors}
\acro{AWGN}{additive white Gaussian noise}
\acro{BB}{base-band}
\acro{BCJR}{Bahl, Cocke, Jelinek and Raviv}
\acro{BER}{bit error rate}
\acro{BFDM}{bi-orthogonal frequency division multiplexing}
\acro{BPSK}{binary phase shift keying}
\acro{BS}{base stations}
\acro{CA}{carrier aggregation}
\acro{CAF}{cyclic autocorrelation function}
\acro{Car-2-x}{car-to-car and car-to-infrastructure communication}
\acro{CAZAC}{constant amplitude zero autocorrelation waveform}
\acro{CB-FMT}{cyclic block filtered multitone}
\acro{CCDF}{complementary cumulative density function}
\acro{CDF}{cumulative density function}
\acro{CDMA}{code-division multiple access}
\acro{CFO}{carrier frequency offset}
\acro{CIR}{channel impulse response}
\acro{CM}{complex multiplication}
\acro{COFDM}{coded-\acs{OFDM}}
\acro{CoMP}{coordinated multi point}
\acro{COQAM}{cyclic OQAM}
\acro{CP}{cyclic prefix}
\acro{CRB}{Cramer-Rao bound}
\acro{CPE}{constant phase error}
\acro{CR}{cognitive radio}
\acro{CRC}{cyclic redundancy check}
\acro{CRLB}{Cram\'{e}r-Rao lower bound}
\acro{CS}{cyclic suffix}
\acro{CSI}{channel state information}
\acro{CSMA}{carrier-sense multiple access}
\acro{CWCU}{component-wise conditionally unbiased}
\acro{D/A}{digital-to-analog}
\acro{D2D}{device-to-device}
\acro{DAC}{digital-to-analog converter}
\acro{DBF}{digital beamforming}
\acro{DC}{direct current}
\acro{DFE}{decision feedback equalizer}
\acro{DFT}{discrete Fourier transform}
\acro{DL}{downlink}
\acro{DMT}{discrete multitone}
\acro{DNN}{deep neural network}
\acro{DSA}{dynamic spectrum access}
\acro{DSL}{digital subscriber line}
\acro{DSP}{digital signal processor}
\acro{DTFT}{discrete-time Fourier transform}
\acro{DVB}{digital video broadcasting}
\acro{DVB-T}{terrestrial digital video broadcasting}
\acro{DWMT}{discrete wavelet multi tone}
\acro{DZT}{discrete Zak transform}
\acro{E2E}{end-to-end}
\acro{eNodeB}{evolved node b base station}
\acro{E-SNR}{effective signal-to-noise ratio}
\acro{EVD}{eigenvalue decomposition}
\acro{FBMC}{filter bank multicarrier}
\acro{FD}{frequency-domain}
\acro{FDA}{frequency diverse array}
\acro{FDD}{frequency-division duplexing}
\acro{FDE}{frequency domain equalization}
\acro{FDM}{frequency division multiplex}
\acro{FDMA}{frequency-division multiple access}
\acro{FEC}{forward error correction}
\acro{FER}{frame error rate}
\acro{FFT}{fast Fourier transform}
\acro{FI}{Fisher information}
\acro{FIR}{finite impulse response}
\acro{FM}		{frequency modulation}
\acro{FMT}{filtered multi tone}
\acro{FO}{frequency offset}
\acro{F-OFDM}{filtered-\acs{OFDM}}
\acro{FPGA}{field programmable gate array}
\acro{FSC}{frequency selective channel}
\acro{FS-OQAM-GFDM}{frequency-shift OQAM-GFDM}
\acro{FT}{Fourier transform}
\acro{FTD}{fractional time delay}
\acro{FTN}{faster-than-Nyquist signaling}
\acro{GFDM}{generalized frequency division multiplexing}
\acro{GFDMA}{generalized frequency division multiple access}
\acro{GMC-CDM}{generalized	multicarrier code-division multiplexing}
\acro{GNSS}{global navigation satellite system}
\acro{GS}{guard symbols}
\acro{GSM}{Groupe Sp\'{e}cial Mobile}
\acro{GUI}{graphical user interface}
\acro{H2H}{human-to-human}
\acro{H2M}{human-to-machine}
\acro{HF}{high frequency}
\acro{HTC}{human type communication}
\acro{I}{in-phase}
\acro{i.i.d.}{independent and identically distributed}
\acro{IB}{in-band}
\acro{IBI}{inter-block interference}
\acro{IC}{interference cancellation}
\acro{ICI}{inter-carrier interference}
\acro{ICT}{information and communication technologies}
\acro{ICV}{information coefficient vector}
\acro{IDFT}{inverse discrete Fourier transform}
\acro{IDMA}{interleave division multiple access}
\acro{IEEE}{institute of electrical and electronics engineers}
\acro{IF}{intermediate frequency}
\acro{IFFT}{inverse fast Fourier transform}
\acro{IoT}{Internet of Things}
\acro{IOTA}{isotropic orthogonal transform algorithm}
\acro{IP}{internet protocole}
\acro{IP-core}{intellectual property core}
\acro{ISDB-T}{terrestrial integrated services digital broadcasting}
\acro{ISDN}{integrated services digital network}
\acro{ISI}{inter-symbol interference}
\acro{ITU}{International Telecommunication Union}
\acro{IUI}{inter-user interference}
\acro{LAN}{local area netwrok}
\acro{LLR}{log-likelihood ratio}
\acro{LMMSE}{linear minimum mean square error}
\acro{LNA}{low noise amplifier}
\acro{LO}{local oscillator}
\acro{LOS}{line-of-sight}
\acro{LoS}{line of sight}
\acro{LP}{low-pass}
\acro{LPF}{low-pass filter}
\acro{LS}{least squares}
\acro{LTE}{long term evolution}
\acro{LTE-A}{LTE-Advanced}
\acro{LTIV}{linear time invariant}
\acro{LTV}{linear time variant}
\acro{LUT}{lookup table}
\acro{M2M}{machine-to-machine}
\acro{MA}{multiple access}
\acro{MAC}{multiple access control}
\acro{MAP}{maximum a posteriori}
\acro{MAE}{mean absolute error}
\acro{MC}{multicarrier}
\acro{MCA}{multicarrier access}
\acro{MCM}{multicarrier modulation}
\acro{MCS}{modulation coding scheme}
\acro{MF}{matched filter}
\acro{MF-SIC}{matched filter with successive interference cancellation}
\acro{MIMO}{multiple-input multiple-output}
\acro{MISO}{multiple-input single-output}
\acro{ML}{maximum likelihood}
\acro{MLD}{maximum likelihood detection}
\acro{MLE}{maximum likelihood estimator}
\acro{MMSE}{minimum mean squared error}
\acro{MRC}{maximum ratio combining}
\acro{MS}{mobile stations}
\acro{MSE}{mean squared error}
\acro{MSK}{Minimum-shift keying}
\acro{MSSS}[MSSS]	{mean-square signal separation}
\acro{MTC}{machine type communication}
\acro{MU}{multi user}
\acro{MVDR}{minimum variance distortionless response}
\acro{MVUE}{minimum variance unbiased estimator}
\acro{NEF}{noise enhancement factor}
\acro{NLOS}{non-line-of-sight}
\acro{NMSE}{normalized mean-squared error}
\acro{NOMA}{non-orthogonal multiple access}
\acro{NPR}{near-perfect reconstruction}
\acro{NRZ}{non-return-to-zero}
\acro{OU}{Ornstein-Uhlenbeck}
\acro{OFDM}{orthogonal frequency division multiplexing}
\acro{OFDMA}{orthogonal frequency division multiple access}
\acro{OMP}{orthogonal matching pursuit}
\acro{OOB}{out-of-band}
\acro{OQAM}{offset quadrature amplitude modulation}
\acro{OQPSK}{offset quadrature phase shift keying}
\acro{OTFS}{orthogonal time frequency space}
\acro{PA}{power amplifier}
\acro{PAM}{pulse amplitude modulation}
\acro{PAPR}{peak-to-average power ratio}
\acro{PB}{pass-band}
\acro{PC-CC}{parallel concatenated convolutional code}
\acro{PCP}{pseudo-circular pre/post-amble}
\acro{PD}{probability of detection}
\acro{pdf}{probability density function}
\acro{PDF}{probability distribution function}
\acro{PDP}{power delay profile}
\acro{PFA}{probability of false alarm}
\acro{PFD}{phase-frequency detector}
\acro{PHY}{physical layer}
\acro{PIC}{parallel interference cancellation}
\acro{PLC}{power line communication}
\acro{PLL}{phase-locked loop}
\acro{PMF}{probability mass function}
\acro{PN}{pseudo noise}
\acro{ppm}{parts per million}
\acro{PRB}{physical resource block}
\acro{PRB}{physical resource block}
\acro{PSD}{power spectral density}
\acro{Q}{quadrature-phase}
\acro{QAM}{quadrature amplitude modulation}
\acro{QoS}{quality of service}
\acro{QPSK}{quadrature phase shift keying}
\acro{R/W}{read-or-write}
\acro{RAM}{random-access memmory}
\acro{RAN}{radio access network}
\acro{RAT}{radio access technologies}
\acro{RC}{raised cosine}
\acro{RF}{radio frequency}
\acro{rms}{root mean square}
\acro{RMSE}{root mean square error}
\acro{RSSI}{received signal strength indicator}
\acro{RRC}{root raised cosine}
\acro{REF}{reference}
\acro{RW}{read-and-write}
\acro{RX}{receiver}
\acro{SC}{single-carrier}
\acro{SCA}{single-carrier access}
\acro{SC-FDE}{single-carrier with frequency domain equalization}
\acro{SC-FDM}{single-carrier frequency division multiplexing}
\acro{SC-FDMA}{single-carrier frequency division multiple access}
\acro{SD}{sphere decoding}
\acro{SDD}{space-division duplexing}
\acro{SDMA}{space division multiple access}
\acro{SDR}{software-defined radio}
\acro{SDW}{software-defined waveform}
\acro{SEFDM}{spectrally efficient frequency division multiplexing}
\acro{SE-FDM}{spectrally efficient frequency division multiplexing}
\acro{SER}{symbol error rate}
\acro{SIC}{successive interference cancellation}
\acro{SINR}{signal-to-interference-plus-noise ratio}
\acro{SIR}{signal-to-interference ratio}
\acro{SISO}{single-input, single-output}
\acro{SMS}{Short Message Service}
\acro{SNR}{signal-to-noise ratio}
\acro{STC}{space-time coding}
\acro{STFT}{short-time Fourier transform}
\acro{STO}{sample-time-offset}
\acro{SU}{single user}
\acro{SVD}{singular value decomposition}
\acro{TX}{transmitter}
\acro{TD}{time-domain}	
\acro{TDD}{time-division duplexing}
\acro{TTD}{true-time delay}
\acro{TDMA}{time-division multiple access}
\acro{TFL}{time-frequency localization}
\acro{TO}{time offset}
\acro{ToA}{time of arrival}
\acro{TDoA}{time difference of arrival}
\acro{TS-OQAM-GFDM}{time-shifted OQAM-GFDM}
\acro{UE}{user equipment}
\acro{UFMC}{universally filtered multicarrier}
\acro{UL}{uplink}
\acro{ULA}{uniform linear array}
\acro{US}{uncorrelated scattering}
\acro{USB}{universal serial bus}
\acro{USRP}{universal software radio peripheral}
\acro{UW}{unique word}
\acro{VLC}{visible light communications}
\acro{VR}{virtual reality}
\acro{VCO}{voltage-controlled oscillator}
\acro{WCP}{windowing and \acs{CP}}	
\acro{WHT}{Walsh-Hadamard transform}
\acro{WiMAX}{worldwide interoperability for microwave access}
\acro{WLAN}{wireless local area network}
\acro{W-OFDM}{windowed-\acs{OFDM}}	
\acro{WOLA}{windowing and overlapping}	
\acro{WSS}{wide-sense stationary}
\acro{ZCT}{Zadoff-Chu transform}
\acro{ZF}{zero-forcing}
\acro{ZMCSCG}{zero-mean circularly-symmetric complex Gaussian}
\acro{ZP}{zero-padding}
\acro{ZT}{zero-tail}
\acro{URLLC}{ultra-reliable low-latency communications}

\acro{HSI}{human system interface}
\acro{HMI}{human machine interface}
\acro{VR} {visual reality} 
\acro{AGV}{automated guided vehicles}
\acro{MEC}{multiaccess edge cloud}
\acro{TI} {tactile Internet}
\acro{IMT}{ international mobile telecommunications}
\acro{GN}{gateway node}
\acro{CN}{control node}
\acro{NC}{network controller}
\acro{SN}{sensor node}
\acro{AN}{actuator node}
\acro{HN}{haptic node}
\acro{TD}{tactile devices}
\acro{SE}{supporting engine}
\acro{AI}{artificial intelligence}
\acro{TSM}{tactile service manager}
\acro{TTI}{transmission time interval}
\acro{NR}{new radio}
\acro{SDN}{software defined networking}
\acro{NFV}{ network function virtualization}
\acro{CPS}{cyber-physical system}
\acro{TSN}{Time-Sensitive Networking}
\acro{FEC}{forward error correction}
\acro{STC}{space-time  coding}
\acro{HARQ}{hybrid automatic repeat request}
\acro{CoMP} {Coordinated multipoint}
\acro{HIS}{human system interface }
\acro{RU}{radio unit}
\acro{CU}{central unit}
\acro{AoD} {angle of departure}
\end{acronym}
    \title{Cramer-Rao Bound for Angle of Arrival Estimates in True-Time-Delay Systems}
	
	\author{
		\IEEEauthorblockN{
			Carl Collmann, Ahmad Nimr, Gerhard Fettweis
			}
			
		\IEEEauthorblockA{
		Vodafone Chair Mobile Communications Systems, Technische Universit\"{a}t Dresden, Germany\\ \small\texttt{\{carl.collmann, ahmad.nimr,  gerhard.fettweis\}@tu-dresden.de}\\
		}
		}
	\maketitle
	\IEEEpeerreviewmaketitle
	
\begin{abstract}
In the context of joint communication and sensing (JC\&S), the challenge of obtaining accurate parameter estimates is of interest.
Parameter estimates, such as the \ac{AoA} can be utilized for solving the initial access problem, interference mitigation, localization of users or monitoring of the environment and synchronization of \ac{MIMO} systems.
Recently, \ac{TTD} systems have gained attention for fast beam training during initial access and mitigation of beam squinting.
This work derives the \ac{CRB} for angle estimates in typical \ac{TTD} systems.
Properties of the \ac{CRB} and the Fisher information are investigated and numerically evaluated.
Finally, methods for angle estimation such as \ac{ML} and established estimators are utilized to solve the angle estimation problem using a uniform linear array. 
\end{abstract}

 
	\acresetall
\section{Introduction}\label{sec:introduction}

In future 6G systems, joint communication and sensing will play an important role by transforming cell infrastructures into perceptive networks \cite{fanliu_9737357}. 
In this context, parameter estimates such as \ac{AoA}/\ac{AoD} derived from sensing are key enablers of smart, context aware mobile communication systems.
These parameter estimates can be used for beam selection, localization, object tracking or synchronization and calibration of increasingly large antenna arrays in \ac{MIMO} systems \cite{rogalin6760595}.

Recently, the use of \acs{TTD} has gained interest for compensation of the beam squint phenomenon \cite{Mewe202509}, which is a significant issue for large relative bandwidths.
The frequency dependent response of \ac{TTD} arrays can also be exploited for fast beam training \cite{Aditya9723402}, \cite{Jans9975036} and angle estimation \cite{2022JSPSy941015B}, \cite{Coll202403}.
The feasibility of utilizing \acs{TTD} for angle estimation \cite{Coll202403} and beam training \cite{Mokri10622779} has been experimentally verified.
For angle estimation in \ac{TTD} systems, various methods such as \ac{OMP}, \ac{ML} based estimation \cite{2022JSPSy941015B}, \ac{LMMSE} \cite{jans9097132} or correlation and model based estimators \cite{Coll202403}.
Previous works have compared the performance of simple model based estimators to common angle estimation methods \cite{Coll202509_2}.
While super-resolution methods such as rootMUSIC provide greater angle estimation accuracy, even in the presence of impairments, simple model based estimators can still provide decent performance with \ac{MAE} $< \SI{1}{\degree}$ \cite{Coll202509_2}.
For fair comparison it should be noted that common angle estimation methods are limited by the assumption that the narrow-band criterion applies, while \ac{TTD} based estimation relies on intentional violation of this criterion\cite{Coll202509_2}.
Other works have extensively investigated the effect of hardware impairments in \ac{TTD} systems, namely time and phase errors \cite{2022JSPSy941015B}.
In \cite{2022JSPSy941015B} a \ac{CRB} is provided, however the bound is derived under nontrivial assumptions about linearization of hardware impairments and relative ratios of their magnitudes.

As the \ac{CRB} can provide a useful method to judge the performance of estimation methods, derivation of the \ac{CRB} for angle estimates in \ac{TTD} systems is of interest.
This work derives an expression for the \ac{CRB} of \ac{AoA} in typical \ac{TTD} systems to provide insights into limitations on angle estimation accuracy.
Common angle estimation methods such as \ac{ML} estimation and model based estimators are compared to the derived \ac{CRB}.

The paper is structured as follows:
Section \ref{sec:system_model} introduces the general system model for a \ac{TTD} SIMO system.
In section \ref{sec:crb} the \ac{CRB} is derived and insights into components of the \ac{FI} are provided.
Section \ref{sec:evaluation} describes the algorithms used for angle estimation and compares their performance with \ac{MSE} as the metric.
The paper is concluded in section \ref{sec:conclusions} with key results.

    \begin{figure}[tb]
    	\centering
    	\includegraphics[width=\linewidth]{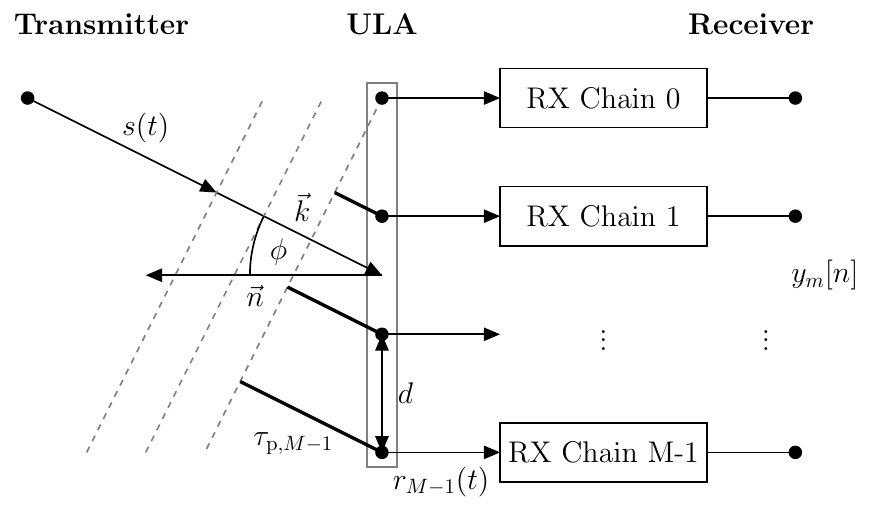}
        \caption{SIMO system model \cite{Coll202509_2} with an omni-directional transmitting antenna and a \ac{ULA} receiver consisting of $M$ elements spaced by distance $d$. The vector $\vec{k}$ denotes the direction from the transmitter (\ac{TX}) to the receiver (\ac{RX}), while the vector $\vec{n}$ represents the array normal.}
    	\label{fig:ula_array}
     \vspace{-5 mm}
\end{figure}

\section{System Model}\label{sec:system_model}
The considered system configuration is illustrated in \figref{fig:ula_array}.
A transmitter, representing a \ac{UE} communicating with a \ac{BS} in uplink, radiates a signal
\begin{align}
    s(t) = \mathfrak{Re} \{ x(t) e^{j2\pi f_\text{c} t}\}
\end{align}
omnidirectionally.
The signal is received by the \ac{BS} equipped with a \ac{ULA} of $M$ elements spaced at $d=\lambda/2$ and with \ac{AoA} $\phi$ relative to the array normal.
This received signal at the $m-$th antenna element with channel gain $\rho$, channel delay $\tau_\text{ch}$ and propagation delay $\tau_{\text{p},m}$ is
\begin{align}
    r_m(t) = \mathfrak{Re} \{ \rho x(t - \tau_\text{ch} - \tau_{\text{p},m}) e^{j2\pi f_\text{c}(t - \tau_\text{ch} - \tau_{\text{p},m})} \}.
\end{align}
After down-conversion the baseband signal delayed by true-time delay $\tau_{\text{d},m}$ is expressed as
\begin{align}
    y_m(t) = \rho x(t - \tau_\text{ch} - \tau_{\text{p},m} - \tau_{\text{d},m}) e^{-j2\pi f_\text{c}(\tau_\text{ch} + \tau_{\text{p},m}) }.
\end{align}
By applying the Fourier transform, the $m$-th signal in the frequency domain is given by
\begin{align}
    Y_m(f) &= \rho X(f) e^{- j2\pi f(\tau_\text{ch} + \tau_{\text{p},m} + \tau_{\text{d},m})} e^{-j2\pi f_\text{c} (\tau_\text{ch} + \tau_{\text{p},m})}\nonumber\\
        &= \rho X(f) e^{-j2\pi (f_\text{c} + f)\tau_\text{ch}} e^{-jm\psi},
\end{align}
with $\psi(f,\phi) = 2\pi[(f_\text{c} + f)\frac{d}{c}\sin\phi + f \tau_{\text{d}}]$.
Then the system function $H=Y/X$ can be expressed as
\begin{align}
    H(f,\phi) = \rho' \sum_{m=0}^{M-1} e^{-jm\psi},
\end{align}
where $\rho'=\rho e^{j\Psi}$ and $\Psi=- 2\pi (f_\text{c} + f)\tau_\text{ch}$.
Using the property of a geometric series, the sum can be rewritten
\begin{align}
    H(f) = \rho' e^{-j\frac{M-1}{2}\psi}\frac{\sin {M\psi/2}}{\sin\psi/2}.
\end{align}
Since for angle estimation a common phase shift applied to the signal is of no significance and the gain of the receive signal can be controlled with \ac{AGC}, for simplicity it is assumed that $\rho'$ is known and compensated for, so that $\rho'=1$.

A noise free observation of the receive signal is 
\begin{align}
    Y(f,\phi) =  {X(f)} \underbrace{\sum_{m=0}^{M-1} e^{-jm2\pi[(f_\text{c} + f)\frac{d}{c}\sin\phi + f \tau_{\text{d}}]}}_{H(f,\phi)},
\end{align}
and noisy observation
\begin{align}
    Z(f) = X(f)H(f,\phi) + V(f),
    \label{eq:sys_model}
\end{align}
with Gaussian noise ${v \sim \mathcal{C}\mathcal{N}(0, \sigma_v^2)}$. 

\begin{figure}[tb]
    	\centering
    	\includegraphics[width=0.9\linewidth]{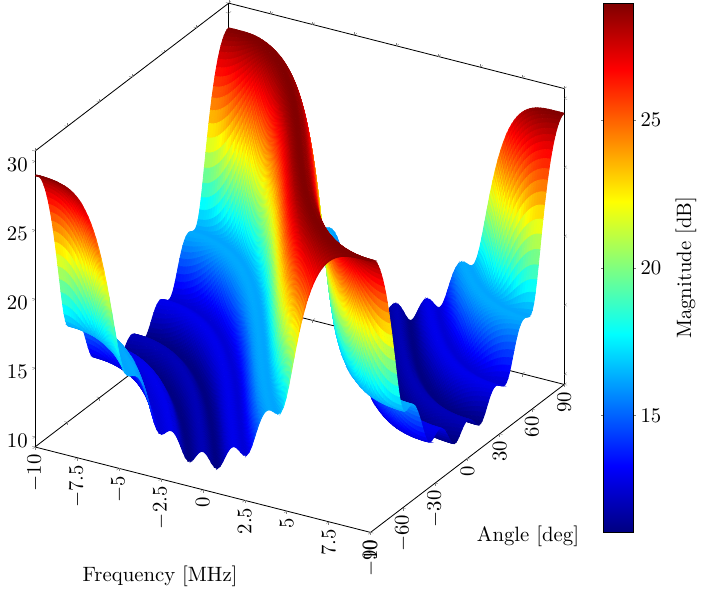}
        \caption{$10\log_\text{10}(\kappa(f,\phi))$ for $M=8,f_\text{c}= \SI{3.75}{GHz}, B=\SI{20}{MHz}, \tau_\text{d}=1/B$}
    	\label{fig:kappa_f_phi}
        \vspace{-5 mm}
\end{figure}
\section{Cramer-Rao Bound for AoA Estimates in TTD Systems}\label{sec:crb}

The objective is to estimate the angle $\phi$ from the received signal based on the previously established model.
Since the Cramer-Rao bound provides the minimum variance that any unbiased estimator can attain and is therefore of interest for evaluating angle estimation methods in \ac{TTD} systems.
When considering $n$ discrete samples/frequency bins at frequencies $f=f_n$, the observation is
\begin{align}
    Z[n] = Y[n,\phi] + V[n] = Z_n.
\end{align}
The likelihood function for the noisy received signal observation vector $\boldsymbol{Z}$ is 
\begin{align}
    p(\boldsymbol{Z};\phi) = \prod_{n=1}^{N} \frac{1}{\pi \sigma_v^2} e^{-\frac{|Z_n - Y_n|^2}{\sigma_v^2}},
\end{align}
and the corresponding log-likelihood function
\begin{align}
    \mathcal{L}(\phi) = \ln p(\boldsymbol{Z};\phi) = -\ln{\pi \sigma_v^2} - \frac{1}{\sigma_v^2}\sum_{n=1}^{N} |Z_n-Y_n|^2.
\end{align}
Taking the first derivative in regards to the phase yields
\begin{align}
    \frac{\partial \mathcal{L}(\phi)}{\partial \phi} &= - \frac{1}{\sigma_v^2}\sum_{n=1}^{N} \left[-\frac{\partial Y_n}{\partial\phi}^*(Z_n - Y_n) - (Z_n - Y_n)^* \frac{\partial Y_n}{\partial \phi}\right]\nonumber\\
    &= \frac{2}{\sigma_v^2}\sum_{n=1}^{N} \mathfrak{Re} \left\{ \frac{\partial Y_n}{\partial\phi}^*(Z_n - Y_n) \right\}.
\end{align}
\begin{figure*}[tb]
    \centering
    \begin{minipage}{0.3\textwidth}
    	\centering
    	\includegraphics[width=\linewidth]{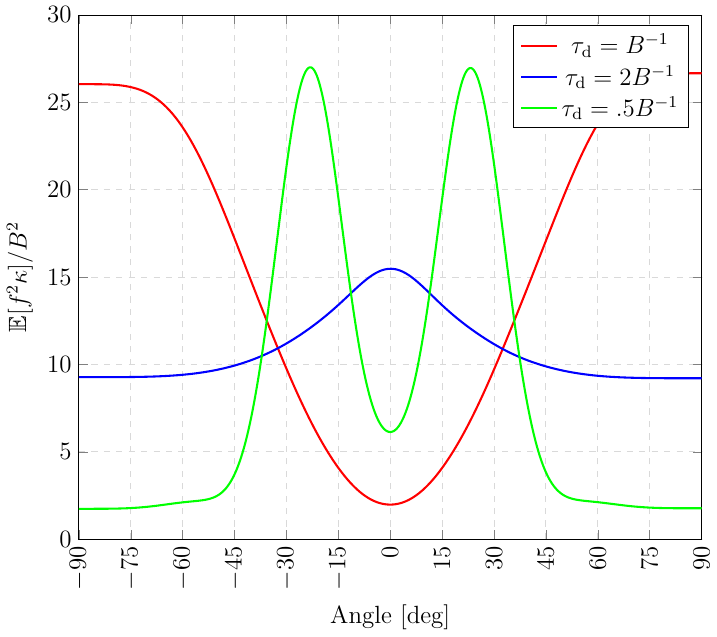}
        \caption{Numerical evaluation of $\kappa_0$ for different $\tau_\text{d}$, $\kappa_0$ is even in regards to $\phi$}
    	\label{fig:kappa_0}
    \end{minipage}
    \begin{minipage}{0.3\textwidth}
    	\centering
    	\includegraphics[width=\linewidth]{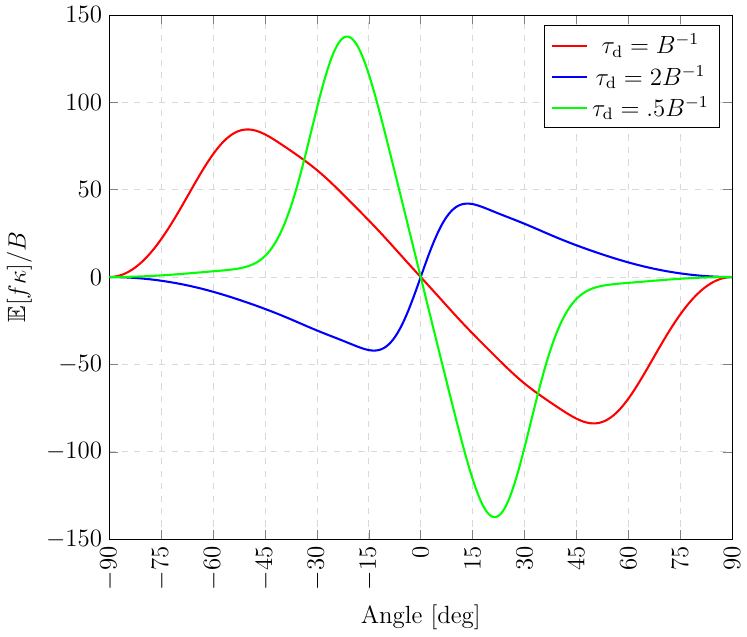}
        \caption{Numerical evaluation of $\kappa_1$ for different $\tau_\text{d}$, $\kappa_1$ is uneven in regards to $\phi$}
    	\label{fig:kappa_1}
    \end{minipage}
    \begin{minipage}{0.3\textwidth}
    	\centering
    	\includegraphics[width=\linewidth]{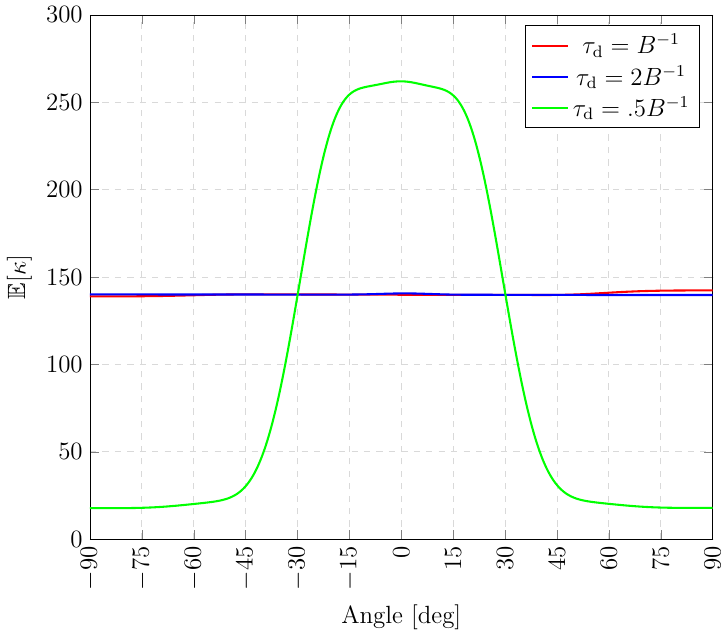}
        \caption{Numerical evaluation of $\kappa_2$ for different $\tau_\text{d}$, $\kappa_2$ is approx. even in regards to $\phi$}
    	\label{fig:kappa_2}
    \end{minipage}
\end{figure*}
The second derivative in regard to the phase is
\begin{align}
    \frac{\partial^2 \mathcal{L}(\phi)}{\partial \phi^2} &= \frac{2}{\sigma_v^2}\sum_{n=1}^{N} \mathfrak{Re} \left\{ \frac{\partial^2 Y_n}{\partial\phi^2}^*(Z_n - Y_n) - \frac{\partial Y_n}{\partial\phi}^* \frac{\partial Y_n}{\partial\phi} \right\}.
\end{align}
The \ac{FI} is defined as
\begin{align}
    &\mathcal{I}(\phi) = - \mathbb{E}\left[ \frac{\partial^2 \mathcal{L}(\phi)}{\partial \phi^2} \right] = \frac{2}{\sigma_v^2}\sum_{n=1}^{N} \left|\frac{\partial Y_n}{\partial\phi} \right|^2\\
    &= \frac{2}{\sigma_v^2}\sum_{n=1}^{N} \left| X_n \left(-j2\pi (f_n + f_\text{c})\frac{d}{c}\cos{\phi}\right) \sum_{m=0}^{M-1} m e^{-j\psi_n} \right|^2\nonumber\\
    &\le \frac{2}{\sigma_v^2}\sum_{n=1}^{N} \left| X_n \right|^2 \cdot \left|2\pi (f_n + f_\text{c})\frac{d}{c}\cos{\phi}\right|^2 \cdot  \underbrace{ \left|\sum_{m=0}^{M-1} m e^{-j\psi_n} \right|^2}_{\kappa_n}.\nonumber
\end{align}
The sum over antenna elements $m$ can be rewritten using the property of a geometric series and taking its first derivative
\begin{align}
    \kappa_n &= \frac{\left| e^{-j\psi_n} - Me^{-jM\psi_n} + (M-1)e^{-j(M+1)\psi_n} \right|^2}{16 \sin^4(\psi_n/2)}\nonumber\\
    &= \frac{(1 - M\cos[(M-1)\psi_n] + (M-1)\cos[M\psi_n])^2}{16 \sin^4(\psi_n/2)}\nonumber\\
    &+\frac{(M\sin[(M-1)\psi_n] - (M-1)\sin[M\psi_n])^2}{16 \sin^4(\psi_n/2)}.
\end{align}
For illustration the variable $\kappa(f,\phi)$ is plotted in \figref{fig:kappa_f_phi}.
It is assumed that frequency bins $f_n$ are taken from the interval $[-B/2, B/2]$ with spacing $\Delta f=B/N$.
The \ac{FI} is rewritten using a Riemann sum
\begin{align}
     \mathcal{I}(\phi) &\le \frac{2}{\sigma_v^2} \left( 2\pi \frac{d}{c} \cos\phi \right)^2\\
     &\cdot\frac{1}{\Delta f} \int_{-B/2}^{B/2} |X(f)|^2 (f+f_\text{c})^2 \kappa (f) df.\nonumber
\end{align}
Under the assumption that the energy of the signal can be written as
\begin{align}
    E_s = \int_{-B/2}^{B/2} |X(f)|^2 df,
\end{align}
and that the spectrum of $X(f)$ is flat and symmetric around $f=0$, the \ac{FI} becomes
\begin{align}
     \mathcal{I}(\phi) &\le \frac{2E_sN}{\sigma_v^2} \left( 2\pi \frac{d}{c} \cos\phi \right)^2 \frac{1}{B} \left[\int_{-B/2}^{B/2} f^2 \kappa (f) df\right.\nonumber\\
     &+\left. 2f_\text{c}\int_{-B/2}^{B/2} f \kappa (f) df + f_\text{c}^2 \int_{-B/2}^{B/2}\kappa (f)df\right].
\end{align}
When treating the frequency as a uniformly distributed random variable $f \sim \mathcal{U}(-B/2,B/2)$, the integrals can be treated as taking the expectation with
\begin{align}
    \mathbb{E}[\kappa] &= \frac{1}{B}\int_{-B/2}^{B/2}\kappa (f)df = \kappa_0,\\
    \mathbb{E}[f\kappa] &= \frac{1}{B}\int_{-B/2}^{B/2}f\kappa (f)df = \kappa_1,\\
    \mathbb{E}[f^2\kappa] &= \frac{1}{B} \int_{-B/2}^{B/2} f^2 \kappa (f) df=\kappa_2.
\end{align}
The expectations are plotted in \figref{fig:kappa_0}, \figref{fig:kappa_1} and \figref{fig:kappa_2} with parameters identical to \figref{fig:kappa_f_phi}.
Inserting into the \ac{FI}
\begin{align}
     &\mathcal{I}(\phi) \le\\
     &\frac{2E_sN}{\sigma_v^2} \left( 2\pi \frac{d}{c} \cos\phi \right)^2  \left[ \mathbb{E}[f^2\kappa] + 2f_\text{c}\mathbb{E}[f\kappa] + f_\text{c}^2 \mathbb{E}[\kappa]\right].\nonumber
    \label{eq:eq_fischer_info_expect}
\end{align}
\begin{figure}[tb]
    	\centering
    	\includegraphics[width=0.8\linewidth]{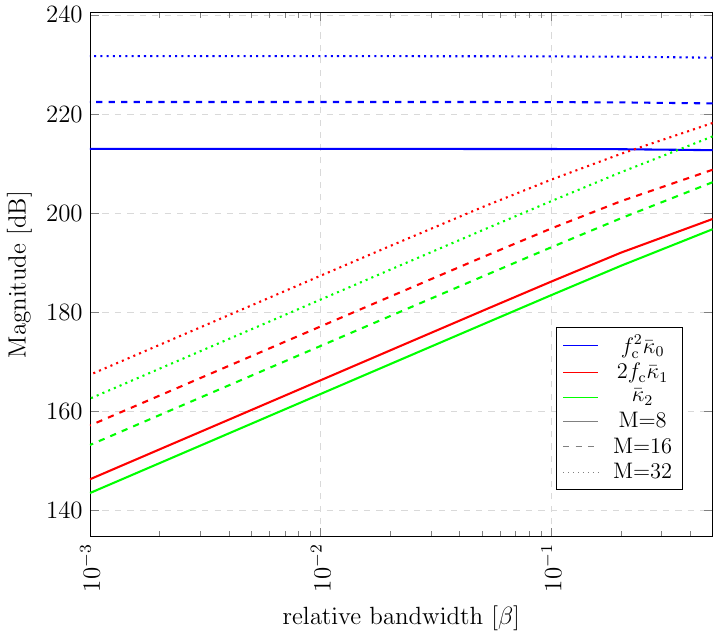}
        \caption{Magnitude in $\SI{}{dB}$ for average $\bar{\kappa}_i$ over relative bandwidth $\beta$ at different number of antennas $M \in \{8,16,32\}$}
    	\label{fig:SNR_MSE_kappa}
\end{figure}
It can be seen that for the narrow-band case of $B=\SI{20}{MHz}$, the term scaled with $f_\text{c}^2$ is dominant.
\figref{fig:SNR_MSE_kappa} shows the average $\bar{\kappa}_i$ over relative bandwidth $\beta=B/f_\text{c}$ at different number of antennas.
For higher relative bandwidths the influence of $\kappa_1,\kappa_2$ is more significant compared to $\kappa_0$.
Under the assumption that $f_\text{c}^2 \gg B^2$ this expression can be further simplified
\begin{align}
    \mathcal{I}(\phi)&\le \frac{2E_sN \kappa_0}{\sigma_v^2} \left( 2\pi f_\text{c} \frac{d}{c} \cos\phi \right)^2.
\end{align}
Then the \ac{CRLB} for $\phi$ is given as
\begin{align}
    \text{Var}(\hat{\phi}) &\ge \frac{1}{\mathcal{I}(\phi)}\nonumber\\
    &\ge \frac{\sigma_v^2}{2E_sN \kappa_0 \left( 2\pi f_\text{c} \frac{d}{c} \cos\phi \right)^2}.\nonumber
    \label{eq:crlb_simpl}
\end{align}

    \section{Evaluation}\label{sec:evaluation}

\subsection{Digital Signal Processing}

\begin{table}[tb]
    \centering
    \begin{center}
        \caption{System parameters.}
        \vspace{-2mm}
            \label{table:system_parameters}
        \resizebox{\linewidth}{!}{
        \begin{tabular}{| l | l | l |} 
         \hline
         Parameter & Symbol & Value \\
         \hline
         Carrier frequency & $f_c$ & $\SI{3.75}{GHz}$ \\ 
         Bandwidth & $B$ & $\SI{20}{MHz}$ \\
         ULA elements & $M$ & $[8, 16, 32]$ \\
         ULA element spacing & $d $ & $ \SI{4}{cm}$\\
         Signal length & $N$ & $512$ samples\\
         True time delay & $\tau_d $ & $ \sim B^{-1}$\\
         Steering angle & $\phi $ & $ [-60 \degree, \hdots, 60 \degree]$\\
         SNR & $\text{SNR}$ & $ [-20, \hdots, 10]~\SI{}{dB}$\\
         \hline
        \end{tabular}
    }
    \end{center}
    \vspace{-5mm}
\end{table}

The observation signal vector $\boldsymbol{Z}$ is generated according to the established system model \eqref{eq:sys_model}.
Signal parameters are given by \tabref{table:system_parameters}, while for this evaluation the narrow-band case $B=\SI{20}{MHz}$ is considered, so that the simplified \ac{CRB} given by \eqref{eq:crlb_simpl} can be used.
Other parameters, such as the carrier frequency are specified in accordance with TU Dresdens 5G campus networks \cite{tu_dd_campusnetworks}.
The evaluation range for the steering angles is limited to the interval $[-60 \degree, \hdots, 60 \degree]$, as outside of this range the \acs{ULA} cant effectively radiate.

For obtaining the \ac{ML} estimate of the \ac{AoA}, the following estimator is used
\begin{align}
    \hat{\phi}_\text{ML} &= \underset{\phi}{\text{argmax}}~ \mathcal{L}(\phi) \\
    & = \underset{\phi}{\text{argmax}} \left[ \sum_{n=1}^{N} |Z_n-Y_n|^2 \right].\nonumber
\end{align}
Note that the noise free system response $\boldsymbol{Y}$ is constructed for a discrete, finite number of angles which imposes a practical limitation on the resolution of this angle estimation approach.
For the peak method, the observation vector is pre-processed by circular convolution with a rectangular window $\boldsymbol{w}$ of size $b$
\begin{align}
    \hat{\boldsymbol{Z}} = \boldsymbol{Z} \circledast \boldsymbol{w}.
\end{align}
The purpose of this pre-processing is to enhance the reliability of finding the frequency component with maximum power 
\begin{align}
    \hat{f} = \underset{f}{\text{argmax}} ~\hat{\boldsymbol{Z}}.
\end{align}
The angle estimate is given by the following function \cite{Coll202403}
\begin{align}
    \hat\phi_\text{TTD}(f=\hat{f}) = \text{arcsin} \left( - \frac{f \tau_d}{f + f_c} \frac{c}{d} \right).
    \label{eq:arcsin_map}
\end{align}

\begin{figure}[tb]
    \centering
    \includegraphics[width=0.8\linewidth]{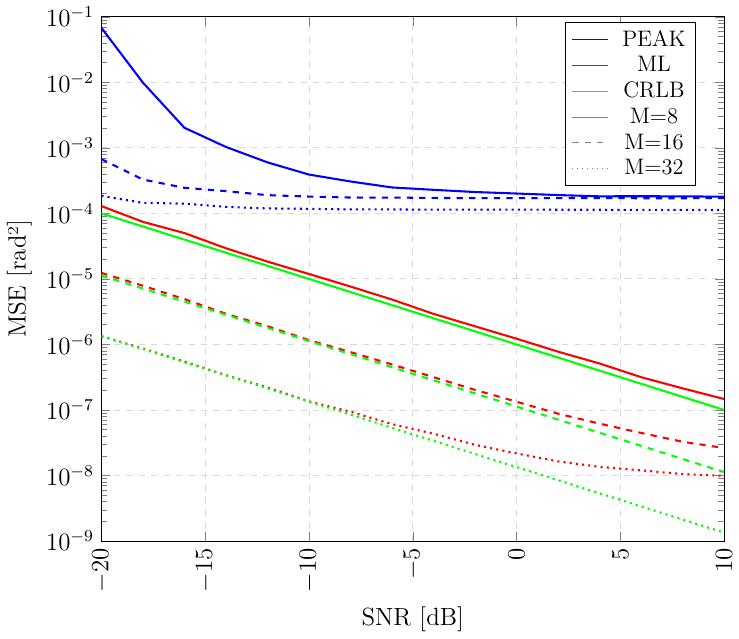}
    \caption{\ac{MSE} for different \ac{SNR}}
    \label{fig:mse_snr}
\end{figure}

\subsection{Results}

\figref{fig:mse_snr} plots the \ac{MSE} for \ac{ML} and peak estimators in combination with the derived \ac{CRB} \eqref{eq:crlb_simpl} over the \ac{SNR} for different number of antennas.
Typically the \ac{CRB} for angle estimates depends on the number of antennas \cite{Kay_Steven10_5555_151045} \cite{trees}.
Note that in \eqref{eq:crlb_simpl}, the dependence of the \ac{CRB} on $M$ is contained in the parameter $\kappa_0$.
As expected, the \ac{CRB} and the estimation performance for both estimators improves with number of antennas.
The estimation performance for the peak method is limited in resolution.
This is explained by the fact that only a limited number of frequency bins is available and that their spacing $N/B$ imposes a limit on achievable \ac{MSE}.
The \ac{ML} estimator closely approaches the \ac{CRB}, while not attaining the \ac{CRB} for higher \ac{SNR}.
While this seems counter intuitive, it is explained by performing the \ac{ML} estimation only with a finite number of angles to construct $\boldsymbol{Y}(\phi)$.
Therefore, the difference between the \ac{ML} estimator and the \ac{CRB} at higher \ac{SNR} can be resolved by simulating $\boldsymbol{Y}(\phi)$ for a higher number of discrete angles.
\section{Conclusion}\label{sec:conclusions}

In this work, the \ac{CRB} for \ac{AoA} estimates in \ac{TTD} systems is derived.
Furthermore, the performance of \ac{ML} and peak based angle estimation methods are compared to the \ac{CRB} with \ac{MSE} as metric.
The \ac{ML} method can attain the \ac{CRB} while the peak based method does not at any \ac{SNR}.
This limitation of the peak method results from only considering the frequency bin with the maximum power for \ac{AoA} estimation.
The spacing between frequency bins imposes a limit on the attainable angle estimation performance of this method.
A potential solution to this problem is to up-sample the observed signal vector with sinc-interpolation to improve the frequency resolution.
For the given parameters, a \ac{RMSE} $<\SI{1}{\degree}$ can be obtained with the peak method, which can be sufficient for solving the initial beam acquisition problem.

Future works should also consider the trade-off between computational complexity and angle estimation performance for the \ac{ML} and peak method.

	\section*{Acknowledgment}

This work was supported by BMBF under the project KOMSENS-6G (16KISK124).
    \bibliographystyle{IEEEtran}
	\bibliography{references}

\end{document}